\documentclass[prl,aps,superscriptaddress,amssymb,twocolumn]{revtex4-1}
\usepackage{graphicx}
\usepackage{amsmath}
\usepackage{latexsym}
\usepackage{dcolumn}
\usepackage{bm}
\usepackage{float}
\usepackage{graphicx,here}

\emergencystretch=\maxdimen
\begin{document}
\title{Discovery of Weyl nodal lines in a single-layer ferromagnet}

\author{Baojie Feng}\thanks{bjfeng@iphy.ac.cn}
\affiliation{Institute of Physics, Chinese Academy of Sciences, Beijing 100190, China}
\affiliation{Hiroshima Synchrotron Radiation Center, Hiroshima University, 2-313 Kagamiyama, Higashi-Hiroshima 739-0046, Japan}
\author{Run-Wu Zhang}
\affiliation{Beijing Key Laboratory of Nanophotonics and Ultrafine Optoelectronic Systems, School of Physics, Beijing Institute of Technology, Beijing 100081, China}
\author{Ya Feng}
\affiliation{Ningbo Institute of Material Technology and Engineering, Chinese Academy of Sciences, Ningbo 315201, China}
\affiliation{Hiroshima Synchrotron Radiation Center, Hiroshima University, 2-313 Kagamiyama, Higashi-Hiroshima 739-0046, Japan}
\author{Botao Fu}
\affiliation{Beijing Key Laboratory of Nanophotonics and Ultrafine Optoelectronic Systems, School of Physics, Beijing Institute of Technology, Beijing 100081, China}
\affiliation{College of Physics and Electronic Engineering, Center for Computational Sciences, Sichuan Normal University, Chengdu, 610068, China}
\author{Shilong Wu}
\affiliation{Graduate School of Science, Hiroshima University, 1-3-1 Kagamiyama, Higashi-Hiroshima 739-8526, Japan}
\author{Koji Miyamoto}
\affiliation{Hiroshima Synchrotron Radiation Center, Hiroshima University, 2-313 Kagamiyama, Higashi-Hiroshima 739-0046, Japan}
\author{Shaolong He}
\affiliation{Ningbo Institute of Material Technology and Engineering, Chinese Academy of Sciences, Ningbo 315201, China}
\author{Lan Chen}
\affiliation{Institute of Physics, Chinese Academy of Sciences, Beijing 100190, China}
\author{Kehui Wu}
\affiliation{Institute of Physics, Chinese Academy of Sciences, Beijing 100190, China}
\author{Kenya Shimada}
\affiliation{Hiroshima Synchrotron Radiation Center, Hiroshima University, 2-313 Kagamiyama, Higashi-Hiroshima 739-0046, Japan}
\author{Taichi Okuda}
\affiliation{Hiroshima Synchrotron Radiation Center, Hiroshima University, 2-313 Kagamiyama, Higashi-Hiroshima 739-0046, Japan}
\author{Yugui Yao}\thanks{ygyao@bit.edu.cn}
\affiliation{Beijing Key Laboratory of Nanophotonics and Ultrafine Optoelectronic Systems, School of Physics, Beijing Institute of Technology, Beijing 100081, China}

\date{\today}

\begin{abstract}

Two-dimensional (2D) materials have attracted great attention and spurred rapid development in both fundamental research and device applications. The search for exotic physical properties, such as magnetic and topological order, in 2D materials could enable the realization of novel quantum devices and is therefore at the forefront of materials science. Here, we report the discovery of two-fold degenerate Weyl nodal lines in a 2D ferromagnetic material, a single-layer gadolinium-silver compound, based on combined angle-resolved photoemission spectroscopy measurements and theoretical calculations. These Weyl nodal lines are symmetry protected and thus robust against external perturbations. The coexistence of magnetic and topological order in a 2D material is likely to inform ongoing efforts to devise and realize novel nanospintronic devices.

\end{abstract}

\maketitle

Spintronics is an emerging technique that uses electron spin as a medium for data storage and transfer\cite{ZuticI2004,JoshiVK2016}. Ferromagnets are widely used as spintronic materials because they possess electronic band structures that are spin split by the exchange interaction. Recently, the desire to miniaturize future quantum devices has stimulated great research interest in low-dimensional materials. For example, various two-dimensional (2D) materials have been realized, such as graphene\cite{Neto2009,AllenMJ2010}, phosphorene\cite{KouL2015,CarvalhoA2016}, and borophene\cite{MannixAJ2015,FengB2016}, providing the possibility to realize novel quantum devices at the atomic scale. The search for 2D ferromagnetic materials is thus a promising route towards future nanospintronics. However, the long-range ferromagnetic order in 2D systems is vulnerable to low-energy spin-wave excitations, making it difficult to realize 2D ferromagnetism; such low-energy excitations can be gapped out by magnetic anisotropy\cite{Siegmann2006}. Intrinsic 2D ferromagnetism has been experimentally observed only recently in several van der Waals crystals, including Cr$_2$Ge$_2$Te$_6$, CrI$_3$, and VSe$_2$\cite{GongC2017,HuangB2017,BonillaM2018}.

The transport properties of a material, which are crucial for its use in spintronic devices, largely depend on the band structure near the Fermi level. Therefore, exploration of ferromagnetic materials with exotic band structures provides great opportunities to realize novel spintronic devices. Recently, topological band structures including the Dirac cone, Weyl cone, and Dirac/Weyl nodal line have attracted great attention because of their potential device applications\cite{WengH2016,FangC2016,YanB2017,ArmitageNP2018}. To date, most materials that host topological band structures are non-magnetic and three dimensional, strongly limiting their applicability to spintronic devices. Therefore, it remains important to search for novel 2D materials with both magnetic order and topological band structure.

Here, we study a novel 2D ferromagnet, single-layer GdAg$_2$, which can be synthesized using a bottom-up approach. Our ab initio molecular dynamic (AIMD) simulation results show that freestanding single-layer GdAg$_2$ is stable at room temperature. Moreover, all the atoms of GdAg$_2$ are coplanar; therefore, its thickness reaches the atomic limit. Recently, Ormaza et al.\cite{OrmazaM2016,CorreaA2016} reported that single-layer GdAg$_2$ grown on Ag(111) is ferromagnetic with a Curie temperature as high as 85 K. Because the electronic structure of a material is critical to understand its physical properties, it is highly desirable to study single-layer GdAg$_2$ by angle-resolved photoemission spectroscopy (ARPES) and first-principles calculations. However, previous works only reported ARPES results along high-symmetry directions\cite{OrmazaM2016}, which is insufficient to understand the electronic structure of single-layer GdAg$_2$. Our detailed ARPES measurements show the existence of Weyl nodal lines that surround the $\Gamma$ point, which agrees well with our theoretical calculations.

\begin{figure*}[htb]
\includegraphics[width=15cm]{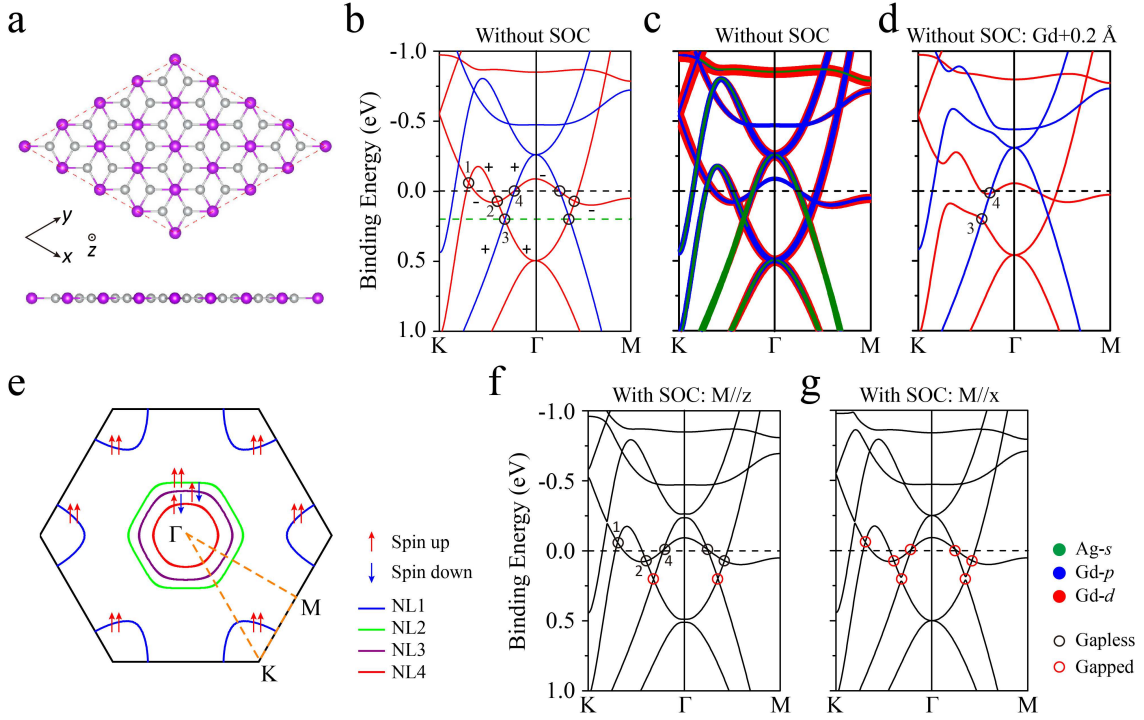}
\caption{First-principles calculation results for freestanding GdAg$_2$. (a) Optimized structure of single-layer GdAg$_2$. Large pink and small grey balls represent Gd and Ag atoms, respectively. (b) Spin-resolved band structures of single-layer GdAg$_2$ without spin--orbit coupling (SOC) effects. Red and blue lines represent spin up and spin down, respectively. The black circles indicate the gapless band crossing points. The plus and minus signs indicate the M$_z$ parity of each band, which are also valid when the SOC is considered. (c) Orbital-projected band structures of single-layer GdAg$_2$ without SOC. (d) Band structures of single-layer GdAg$_2$ after shifting the Gd atoms by 0.2 {\AA} in the out-of-plane direction. (e) Momentum distribution of the band crossing points without SOC. The crossing points form four nodal lines: NL1, NL2, NL3, and NL4. (F) and (G) Calculated band structures with SOC. The magnetization direction is out of plane in (f) and in plane in (g). The gapless and gapped band crossing points are indicated by black and red circles, respectively.}
\end{figure*}

Figure 1a shows the relaxed atomic structure of freestanding single-layer GdAg$_2$, which is composed of a honeycomb Ag lattice and triangular Gd lattice. All the atoms are coplanar, and thus the mirror reflection (M$_z$) symmetry is respected  when the electron spin is neglected. Our AIMD simulations confirmed the stability of freestanding GdAg$_2$ at room temperature (Supplementary Note 1). Figure 1b shows the band structures of GdAg$_2$ without considering spin--orbit coupling (SOC). All the bands are spin polarized because of the presence of half-filled Gd 4f orbitals . Near the Fermi level, there are four band crossings along the $\Gamma$--K direction and three along the $\Gamma$--M direction, as indicated by black circles. These bands are primarily derived from Ag s, Gd d, and Gd p orbitals (Fig. 1c). In Fig. 1E, we present the momentum distribution of the band crossing points. The crossing points form four nodal lines in the Brillouin zone: one surrounds the K (K') point and three surround the $\Gamma$ point; hereafter, we label these four nodal lines NL1, NL2, NL3, and NL4, respectively.

\begin{figure}[htb]
\includegraphics[width=8.5cm]{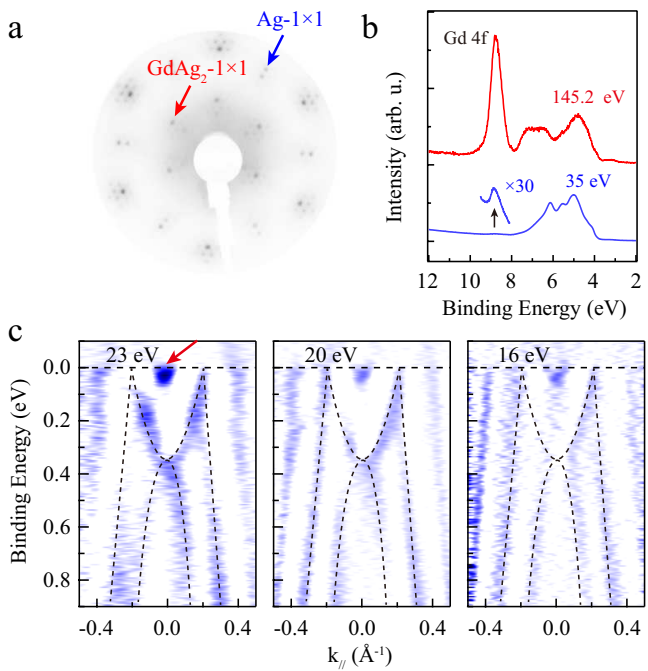}
\caption{Low-energy electron diffraction (LEED) and photoemission measurements of single-layer GdAg$_2$ on Ag(111). (a) A LEED pattern of single-layer GdAg$_2$ on Ag(111) measured with a beam energy of 50 eV. The 1$\times$1 spots of Ag(111) and GdAg$_2$ are indicated by blue and red arrows, respectively. (b) Angle-integrated photoemission spectra of single-layer GdAg$_2$ on Ag(111), which include both on (red) and off (blue) resonance of the Gd 4f levels. The spectra were acquired with photon energies of 145.2 eV (red) and 35 eV (blue). (c) Second-derivative ARPES intensity images along the $\Gamma$--M direction acquired with photon energies of 16, 20, and 23 eV. White dashed lines are a visual guide to trace the band dispersions.}
\end{figure}

Now that we have identified the existence of several nodal lines in single-layer GdAg$_2$, an interesting question is whether these nodal lines are symmetry protected. In Fig. 1b, we show the M$_z$ parity of each band by a plus or minus sign. For NL1, NL2, and NL4, the two crossing bands have opposite M$_z$ parities, which indicates there is no band hybridization when the two bands intersect. Therefore, NL1, NL2, and NL4 are protected by the M$_z$ symmetry. For NL3 and NL4, the two crossing bands have opposite spin polarization. In the absence of SOC, the spin is a good quantum number, and thus the spin degeneracy of NL3 and NL4 is intrinsic for single-layer GdAg$_2$. As a result, NL3 and NL4 cannot be gapped out by weak perturbations, even if the crystal symmetry is broken. To test the robustness of NL3 and NL4, we slightly moved the Gd atoms in the out-of-plane direction, which broke the M$_z$ symmetry. The calculated band structures are shown in Fig. 1D. As expected, NL1 and NL2 disappeared because the M$_z$ symmetry was broken, whereas NL3 and NL4 survived.

When the SOC is considered, the spin is no longer a good quantum number, and the degeneracy of NL3 will be lifted because of the band hybridization. Because the SOC in GdAg$_2$ is rather weak, the energy gap is negligible and therefore NL3 is approximately preserved. As discussed above, the M$_z$ symmetry is always respected without SOC. However, in the presence of SOC, the M$_z$ symmetry will be broken when the magnetization direction is in plane; this will gap out all the nodal lines. Conversely, when the magnetization direction is out of plane (M//z), NL1, NL2, and NL4 will survive because the M$_z$ symmetry is not broken. Our calculation results agree well with this analysis, as illustrated in Figs. 1f and 1g. These results show that it is possible to switch the gap on/off by manipulating the magnetization direction.

To experimentally confirm this exotic electronic structure of single-layer GdAg$_2$, we prepared single-layer GdAg$_2$ on an Ag(111) surface and conducted photoemission spectroscopy measurements. A low-energy electron diffraction (LEED) pattern of GdAg$_2$/Ag(111) is shown in Fig. 2a. The 1$\times$1 spots of Ag(111) and GdAg$_2$ are indicated by blue and red arrows, respectively.  Because of the lattice mismatch, there is a moir\'{e} pattern, which is 12$\times$12 with respect to the Ag(111)-1$\times$1 lattice, in agreement with a previous report\cite{OrmazaM2016}. The formation of a moir\'{e} pattern indicates the relative stability of freestanding single-layer GdAg$_2$. Figure 2b displays integrated photoemission spectra of GdAg$_2$/Ag(111) measured with different photon energies. The occupied Gd 4f peak is located 8.8 eV below the Fermi level, which is slightly deeper than that in a Gd single crystal\cite{FrietschB2015}. It should be noted that the intensity of the Gd 4f peak at a photon energy of 145.2 eV was markedly enhanced, which arises from the Gd 4d$\rightarrow$4f resonance. The unoccupied Gd 4f peak was not visible in our measurements because it is located above the Fermi level\cite{FausterTh1984}. These results indicate the high quality of our sample.

An ARPES intensity image along the $\Gamma$--K direction is shown in Fig. 2c. We observed remnant  signals from the Shockley surface states of Ag(111), as indicated by the red arrow. This is because the coverage of GdAg$_2$ was less than one monolayer so there were regions of bare Ag(111) surface exposed to the excitation photons. Besides the surface states of Ag(111), we also observed band structure from the surface GdAg$_2$ layer, as indicated by the white dashed lines. These experimental findings agree well with our calculation results for freestanding GdAg$_2$, except that the experimental Fermi level is 0.2 eV lower than the theoretical one, as indicated by the green dashed line in Fig. 1b. Consequently, we observed the lower part of NL3; the upper part was above the Fermi level and thus invisible in our ARPES measurements. The experimental band structure remained unchanged when we varied the energy of incident photons, which confirmed their 2D nature.

\begin{figure*}[htb]
\includegraphics[width=15cm]{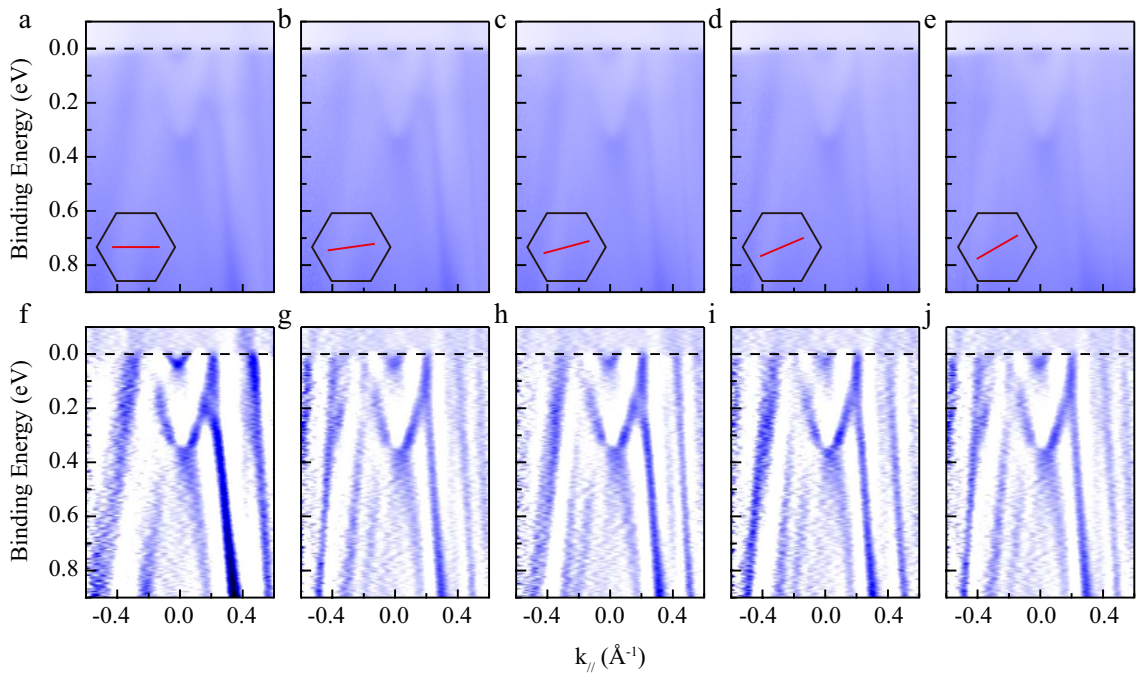}
\caption{ARPES measurements of the electronic structure of single-layer GdAg$_2$ on Ag(111). (a--b) ARPES intensity images along various momentum cuts from $\Gamma$--K to $\Gamma$--M. The corresponding momentum paths in the Brillouin zone are shown in the inset. The photon energy is 21 eV and the polarization is {\it p}. (f--j) Corresponding second-derivative images to enhance the contrast of the bands.}
\end{figure*}

Figure 3a--e show ARPES intensity images along different azimuth angles from $\Gamma$--K to $\Gamma$--M. The corresponding second-derivative images are shown in Fig. 3f--j. We find that the band crossing at the Fermi level, i.e., NL3, does not have a gap along each cut. These results indicate that the band crossing points near the Fermi level form a closed line surrounding the $\Gamma$ point, which agrees well with the shape of NL3. From our ARPES results, the Fermi velocity of NL3 is as high as 5.6 eV$\cdot$\AA, which is comparable to that of graphene\cite{Neto2009,AllenMJ2010}.

Because the Ag(111) substrate is metallic, it tends to hybridize with the bands of the surface GdAg$_2$ layer, which could destroy the band structure of pristine GdAg$_2$. However, the similarity between our experimental results and theoretical calculations suggests that NL3 survives despite the band hybridization. To further study the band hybridization in GdAg$_2$/Ag(111), we performed first-principles calculations including the Ag(111) substrate. Typically, there are two different adsorption geometries of single GdAg$_2$ on Ag(111) in which the Gd atoms sit on the fcc or hcp  sites of the underlying Ag(111) substrate. The calculated band structures of these two geometries are identical, as shown in Fig. 4. As expected, we find that NL3 identified for free-standing GdAg$_2$ survives in GdAg$_2$/Ag(111), as highlighted by the black circles. Orbital composition analysis revealed that these bands are mainly derived from Ag s, Gd p, and Gd d orbitals, which is in excellent agreement with the orbital composition of NL3 in the free-standing layer (Fig. 1C). Although NL3 survived, the other nodal lines in free-standing GdAg$_2$ (NL1, NL2, and NL4), disappeared because of the band hybridization.

\begin{figure}[htb]
\includegraphics[width=8.5cm]{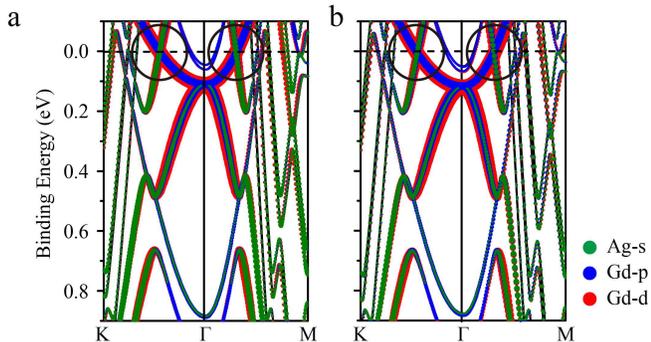}
\caption{Orbital-projected band structures of single-layer GdAg$_2$ on Ag(111). The Gd atoms sit on the (a) fcc and (b) hcp  sites of the underlying Ag(111) substrates. The calculated Fermi level has been shifted by 0.75 eV to achieve better agreement with the experimental results. Contributions from the Ag s, Gd p, and Gd d orbitals are shown in green, blue, and red, respectively. Black circles highlight the position of NL3.}
\end{figure}

In single-layer GdAg$_2$, the localized Gd 4f orbitals  are half filled and contribute to the atomic moments with 7$\mu\rm_B$ per atom; such large magnetic moments will spin split the electronic bands near the Fermi level. According to our calculations, the splitting of the bands is approximately 0.75 eV, which is comparable to that in single-crystal Gd\cite{FrietschB2015}. Because the 4f electrons have no spatial overlap, the neighboring 4f moments align via Ruderman--Kittel--Kasuya--Yosida (RKKY) exchange coupling with conduction electrons, including the Gd p, Gd d, and Ag s orbitals. In other words, the RKKY interaction is mediated by conduction electrons near the Fermi level. In a metallic system, the electronic structure near the Fermi level can usually be treated as  free electrons, and the RKKY coupling shows oscillatory behavior with alternating ferromagnetic and antiferromagnetic coupling depending on the distance between  neighboring magnetic atoms\cite{Siegmann2006}. However, in topological materials, the low-energy excitations behave as Dirac, Weyl, or Dirac/Weyl nodal-line fermions. An RKKY interaction mediated by Dirac or Weyl fermions will show unique behavior, and has attracted great attention recently\cite{LiuQ2009,Black2010,ChangHR2015}. Our discovery of Weyl nodal lines in single-layer GdAg$_2$ could stimulate further research effort on RKKY interactions mediated by nodal-line fermions in the 2D limit.

In summary, our experimental and theoretical results confirmed the existence of symmetry-protected Weyl nodal lines in single-layer GdAg$_2$. In particular, one nodal line is located at the Fermi level when GdAg$_2$ is grown on Ag(111). These results could spur further study on the interplay of ferromagnetic and topological order in the 2D limit. Unlike other 2D ferromagnetic materials that have been discovered recently\cite{GongC2017,HuangB2017}, the thickness of single-layer GdAg$_2$ reaches the atomic limit, which could enable the fabrication of quantum devices at a smaller scale. Because our AIMD simulations indicated the stability of freestanding GdAg$_2$, we expect that single-layer GdAg$_2$ could be epitaxially grown on other substrates or even peeled off from a substrate.

\begin{acknowledgments}

{\center\bf Acknowledgments \par}

The ARPES measurements were performed with the approval of the Proposal Assessing Committee of Hiroshima Synchrotron Radiation Center (Proposal No. 17AG011 and 17BG016). B. Feng was supported by the Japan Society for the Promotion of Science (JSPS) Grant-in-Aid for Early-Career Scientists (Grant No. 18K14191). Y. Yao was supported by the National Key R\&D Program of China (Grant No. 2016YFA0300600) and the NSF of China (Grant No. 11734003 and 11574029). Author contributions: B.Feng conceived the research. B.Feng, Y.F., S.W., K.M., and T.O. performed the ARPES measurements. R.W.Z., B.Fu, and Y.Y. performed the first-principles calculations. All authors contributed to the discussion of the manuscript.

Baojie Feng, Run-Wu Zhang, and Ya Feng contributed equally to this work.

\end{acknowledgments}

\end{document}